\documentstyle[11pt,newpasp,twoside,epsf]{article}
\begin{document}
\title{Gas, Stars and Baryons in Low Surface Brightness Galaxies}
\author{K. O'Neil}
\affil{Arecibo Observatory; HC3 Box 53995;Arecibo, Puerto~Rico; 00612; {koneil@naic.edu}}
\begin{abstract}
Recent surveys have discovered hundreds of low surface brightness galaxies
in the local (z $<$ 0.1) Universe.  Plots of the surface brightness distribution
(the space density of galaxies plotted against central surface brightness)
show a flat distribution from the bright-end cutoff of 21.65 through the current
observational limit of 25.0 B mag arcsec$^{-2}$.  As {\it no} trend is seen to indicate the
size or mass of galaxies decreases with decreasing central surface brightness,
it is likely that a significant percentage of the baryon content in the universe
is contained in these diffuse systems.  In this paper I briefly review the
known properties of low surface brightness galaxies, and describe some current theories
on the baryonic mass fraction of low surface brightness systems and their consequences.
\end{abstract}

\section{Why Study Low Surface Brightness Galaxies?}

Recent surveys by O'Neil, et al. (1997a, 1997b, 2000, 2001)
have discovered hundreds of low surface brightness (LSB) galaxies
in the local universe.  Plots of the surface brightness distribution --
that is, the space density of galaxies plotted against central surface
brightness -- show a flat distribution from  
21.65 through the current observational limit of
25.0 B mag arcsec$^{-2}$ (Figure 1a).  LSB systems therefore numerically
dominate the galaxy population of the local universe.  Additionally,
as no trend is seen to indicate the size or mass of galaxies 
decreases with decreasing central surface brightness, it is 
likely that
a significant percentage of the baryon content in the universe
is contained in these diffuse systems.

\begin{figure}[t]
\begin{center}
\plottwo{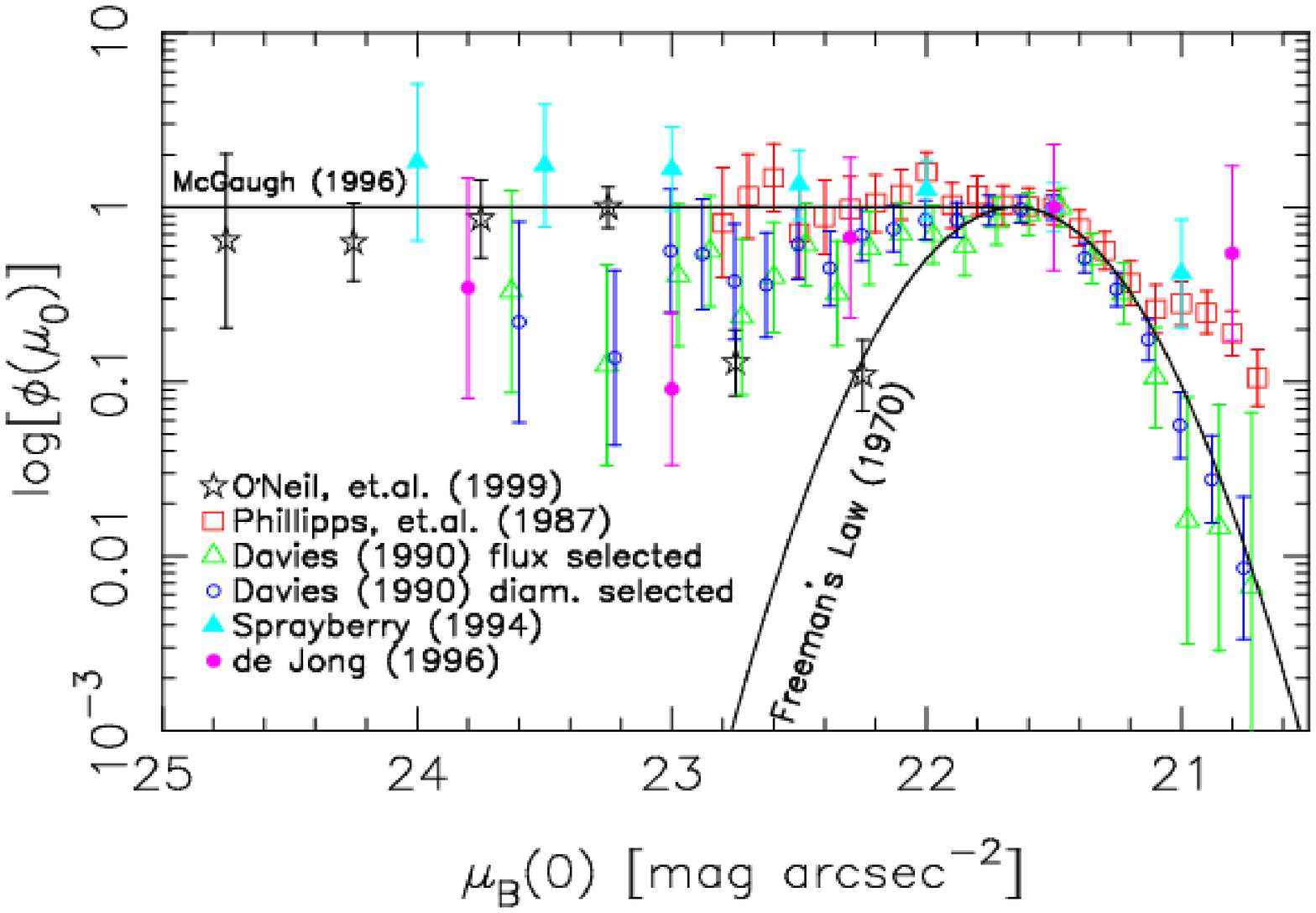}{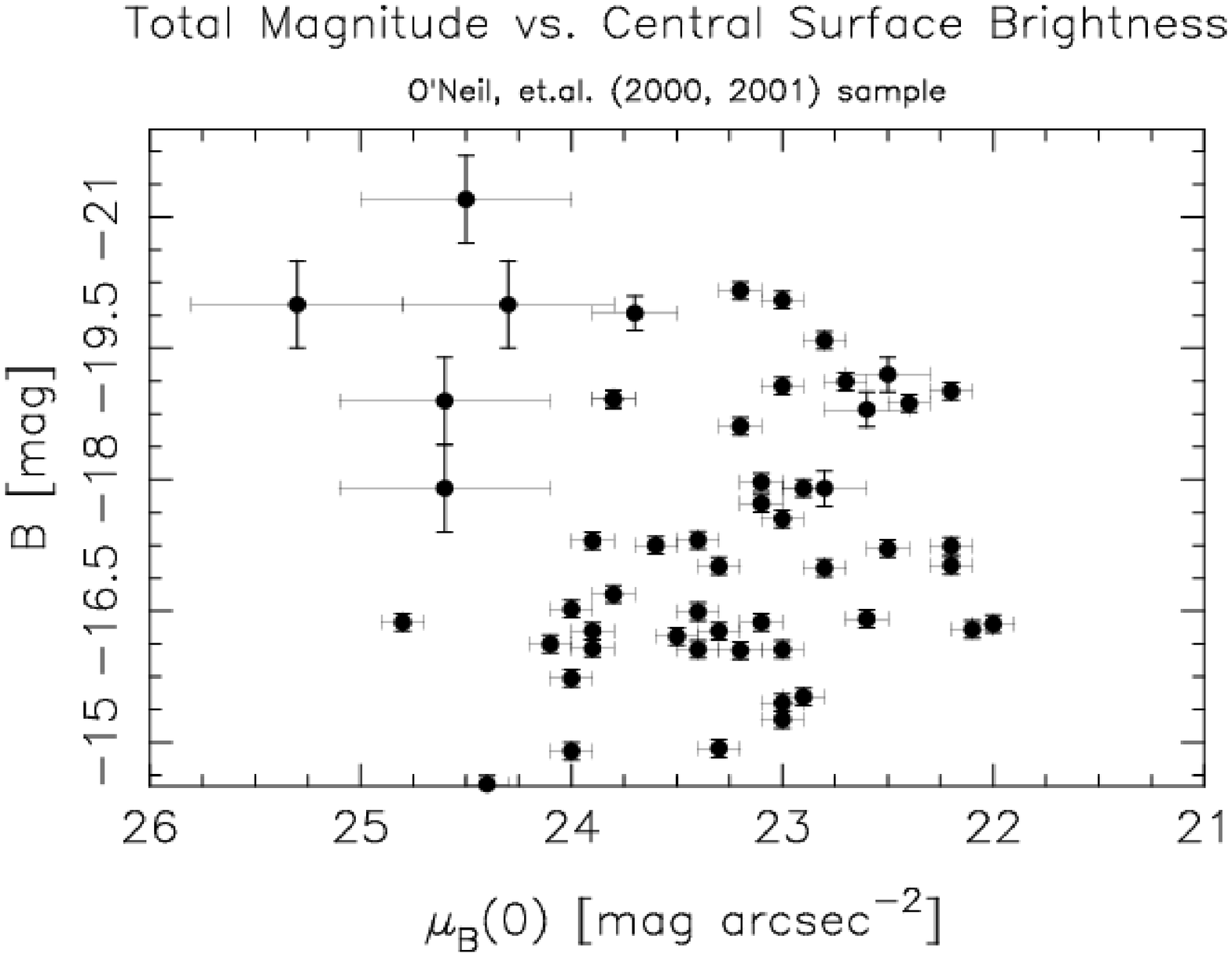}
\caption{(a) The number density of galaxies in the local Universe,
with $\phi$ normalized to 1.  (b) Total magnitude versus  central
surface brightness for a sample of LSB galaxies. }
\end{center}
\end{figure}

\section{LSB Galaxy Sizes}

Contrary to a the often held belief that LSB and dwarf galaxies
are synonymous, no correlation is seen between the central
surface brightness of the galaxies in the O'Neil, {\it et.al.}
samples and either total magnitude or
total (gas) mass (Figure 1b).  Galaxies do not become preferentially
smaller with lower surface brightness.  Instead,  LSB galaxies occupy the 
same luminosity space as their HSB counterparts.

\section{The Mass-to-Luminosity Ratio of LSB Galaxies}

A subset of the O'Neil, {\it et.al.} (1997a, 1997b, 2000, 2001)
LSB galaxies have rotational velocities
$\geq$ 200 km s$^{-1}$ and total luminosities at least an order
of magnitude below L$_*$.  As such they represent extreme
departures from the standard Tully-Fisher relation.
In fact, the sample does not appear to have any significant
correlation between velocity widths and absolute magnitudes,
with only 40\% of the galaxies falling within the 1$\sigma$ LSB
Tully-Fisher relation (Figure 2a). Unless the  percentage of dark matter in
these systems is unusually high, this may indicate the galaxies do not
lie in the same evolutionary state as galaxies with lower
gas content.  Another possible interpretation, though, is found
by thinking of the Tully-Fisher relation as a baryonic versus total
mass relationship (i.e. McGaugh, {\it et.al.} 2000).   In this case it 
can be seen that putting the LSB galaxies onto the baryonic Tully-Fisher
relation is akin to increasing the galaxies baryonic mass-to-light
ratios (Figure 2b).

Additional support for increasing the baryonic mass-to-light
of LSB galaxies can be found by looking at the  rotation curves
of LSB galaxies.  In a recent paper, Swaters, {\it et.al} (2000)
has shown that the baryonic to dark matter ratio ($M_b/M_{DM}$) of LSB galaxies
can be made to mimic to that of HSB galaxies (as opposed to being dark matter
dominated even in the central regions) if $M_b/L$ is allowed
to range from 1$M_{\odot}/L_{\odot}$ through 10$M_{\odot}/L_{\odot}$ (or even higher).

One method for increasing $M_b/L$ in LSB galaxies is to allow the
galaxies to have an initial mass function (IMF) which produces
primarily low mass stars.  This can be brought about
by assuming the low density inherent in LSB systems (often well below
the Kennicutt criterion for star formation, i.e. van Zee, {\it et.al}
1998; de Blok, McGaugh, \& van der Hulst 1996) affects the IMF in
such a way as to prevent large scale production of stars with mass greater than
$2M_{\odot}$.  In addition to putting the LSB systems back on the baryonic
Tully-Fisher relation, a
preferentially low mass IMF can also explain the red, gas-rich LSB
galaxies found by O'Neil, Bothun \& Schombert (2000) as well
as the non-detection of significant numbers of red giant stars in a HST
study of three nearby dE LSB systems by O'Neil, Bothun \& Impey (1999).


\begin{figure}[t]
\begin{center}
\plottwo{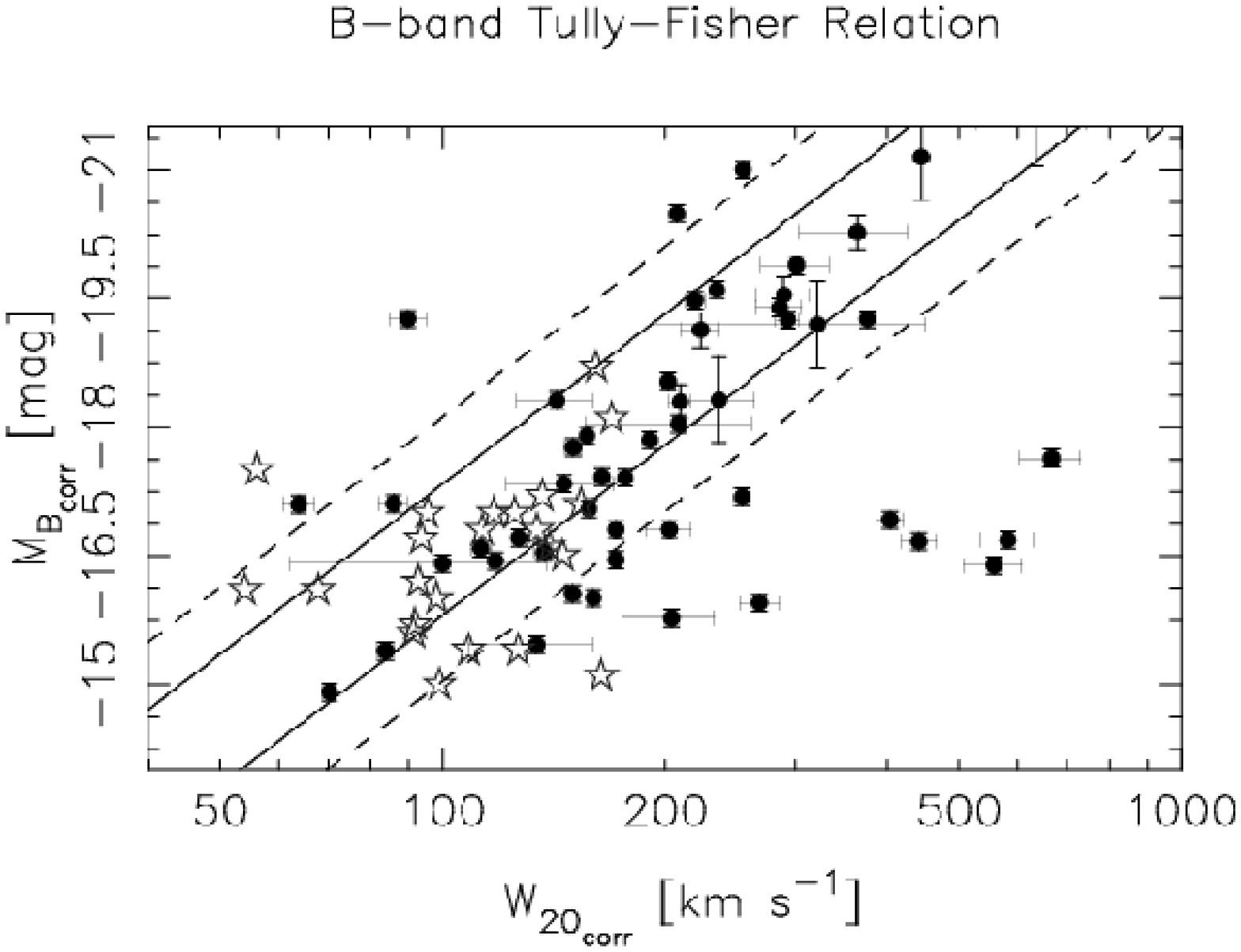}{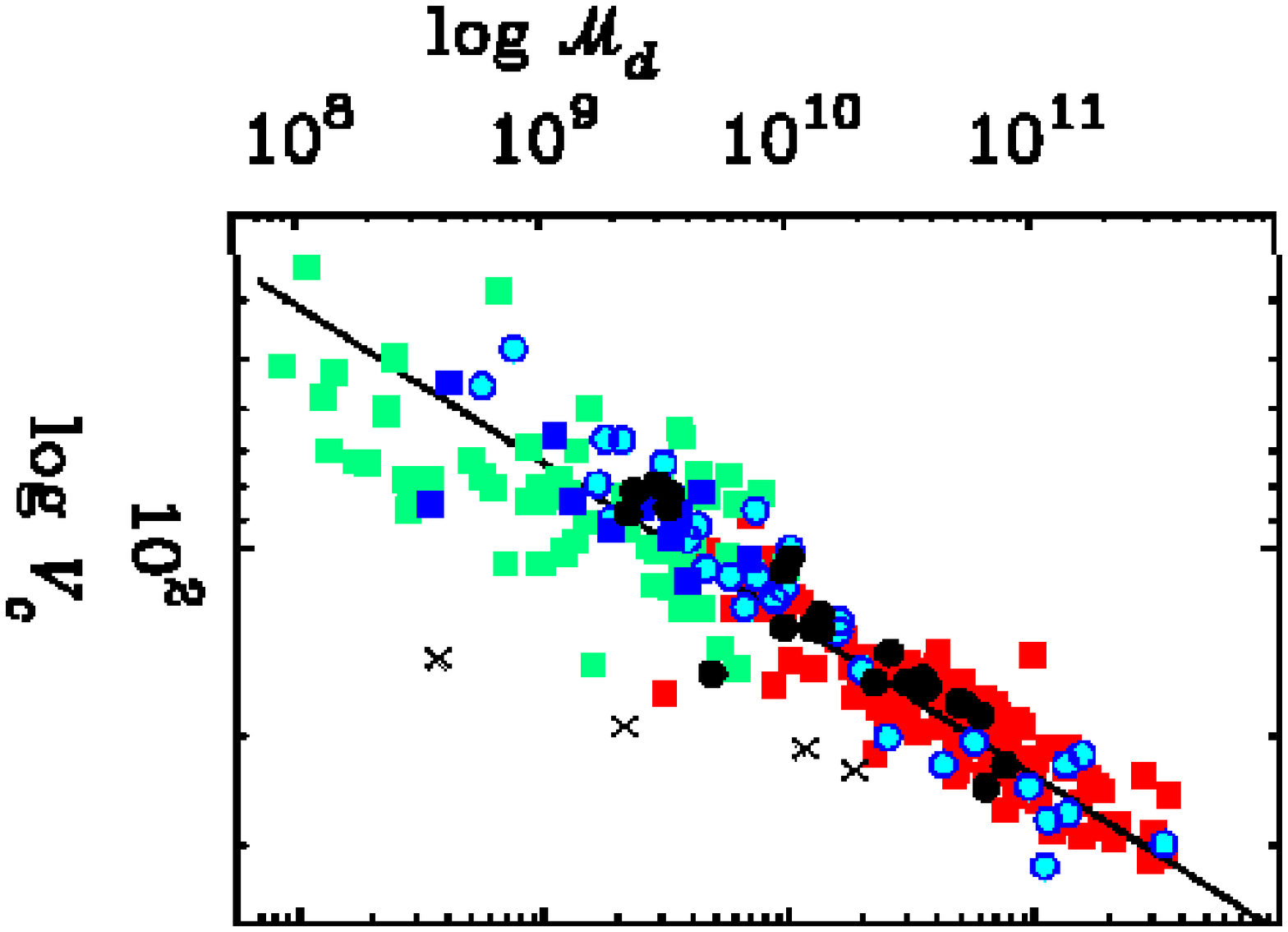}
\caption{(a) A Tully-Fisher plot for a sample of LSB (circles) and late-type (stars)
galaxies.  The solid and dashed lines are the 1$\sigma$ and 2$\sigma$ fits
to the Tully-Fisher relation of Zwaan, {\it et.al} (1995).  The LSB galaxy data is
from O'Neil, Bothun, \& Schombert (2000) and the late-type galaxies are from
Matthews \& Gallagher (1998). (b) The Tully-Fisher relation plotted as
baryonic disk mass versus rotational velocity, from McGaugh, {\it et.al} (2000).
The crosses represent four of the outlying galaxies from the LSB galaxy
Tully-Fisher relation shown in (a).}
\end{center}
\end{figure}

\section{The (Potential) Baryonic Contribution of LSB Galaxies}
It was shown (above) that the number density of LSB galaxies is at least equal to
that of HSB galaxies, and that the mass-to-luminosity ratio of LSB galaxies
is the same, or potentially much higher than, their HSB counterparts.  With these
two ideas in mind, then, we can do a rough calculation of the baryonic contribution
of LSB galaxies to the local universe.  (All the estimates for baryon density
are adapted from are from Fukugita, Hogan, \& Peebles 1998).

First, assume $ L\times{{M}\over{L}}$ remains constant with decreasing central
surface brightness.  Then assume the number density of galaxies ($\phi[\mu_B(0)]$)
is constant out to  26.0 B mag arcsec$^{-2}$, where it cuts off (i.e. we have seen
at least a sampling of all the galaxies in the local Universe).  In this case,
LSB galaxies contribute 9 times the baryon density of HSB galaxies.
The total contribution of disk and irregular galaxies ($\rho_{disk\:+Irr}$)
to the local baryon density is then $0.0084\:h_{70}^{-1}$, and LSB galaxies potentially
contribute 40\% all baryons in the local Universe.

We can now retain the assumption that $ L\times{{M}\over{L}}$ remains constant with decreasing
central surface brightness, but assume $\phi[\mu_B(0)]$ does not cut-off until
30.0 B mag arcsec$^{-2}$.  Here, $\rho_{LSB}\: =\:17\times\rho_{HSB}$,
$\rho_{disk\:+Irr}$ = $0.016\:h_{70}^{-1}$,
and LSB galaxies could contribute 
75\% all baryons in the local Universe.

Finally, we must recall the arguments if the last two sections, and consider the case where
$ L\times{{M}\over{L}}$ increases to 6 times its value between 22.0  B mag arcsec$^{-2}$
and 24.0  B mag arcsec$^{-2}$, at which point it again remains constant.
To be conservative, again assume that we have seen a sampling of all the
galaxies in the Universe (i.e. $\phi[\mu_B(0)]$ cuts-off at 26.0 B mag arcsec$^{-2}$).
For this scenario, $\rho_{LSB}\: =\:22\times\rho_{HSB}$, $\rho_{disk\:+Irr}$ = $0.020\:h_{70}^{-1}$,
and LSB galaxies could contribute
97\% all baryons in the local Universe.  It is now a trivial step to allow
$\phi[\mu_B(0)]$ to remain constant only to  26.5 B mag arcsec$^{-2}$,
and thereby show that LSB galaxies contribute 100\% of all the baryons in the
local Universe.
 
Clearly something is wrong.  Recent studies (i.e. Fukugita, Hogan, \& Peebles 1998)
have shown that all the baryons previously perceived to be `missing' from the local
Universe can be found in the form of ionized gas.  Yet their accounting assumed
LSB galaxies to be low mass objects, an assumption which has since been shown to be incorrect.
We are now faced with the dilemma that instead of our observational counts
showing an under density of baryons in the local Universe when compared with
theoretical predictions, we have a clear over density.  At this point all
assumptions -- in the theoretical models, in determining the ionized gas
contribution, and in determining the space density and baryonic content
of LSB galaxies -- must be re-evaluated and re-tested, so that an accurate
picture of the local baryonic content can be determined.

\acknowledgements
Thanks to Stacy McGaugh for providing Figure 3b, and to
D\&B for their considerable help.

\end{document}